\documentclass{jps-cp}
\usepackage{txfonts}

\usepackage{bbm}
\usepackage{color}

\title{
  Effects of Medium Modifications of Nucleon Form
  Factors on Neutrino Scattering in Dense Matter}

\author{Parada~T.~P.~\textsc{Hutauruk}$^{1}$,
  Yongseok~\textsc{Oh}$^{2,1}$, and Kazuo~\textsc{Tsushima}$^{3,1}$}

\inst{$^{1}$Asia Pacific Center for Theoretical Physics, Pohang, Gyeongbuk 37673, Korea \\
  $^{2}$Department of Physics, Kyungpook National University, Daegu 41566, Korea\\
  $^{3}$Laborat\'{o}rio de F\'{i}sica Te\'{o}rica e Computacional-LFTC, Universidade Cruzeiro
  do Sul, 01506-000, S\~ao Paulo, SP, Brazil}

\email{parada.hutauruk@apctp.org}

\recdate{January 18, 2019}

\abst{Effects of the in-medium modifications of nucleon form factors on
  neutrino interaction in dense matter are presented by considering both the weak and 
  electromagnetic interactions of neutrinos with the constituents of the matter. 
  A relativistic mean field and the quark-meson coupling models are respectively 
  adopted for the effective nucleon mass and in-medium nucleon form factors.
  We calculate the cross section of neutrino scattering as well as the neutrino mean
  free path. We found the cross sections of neutrino scattering in cold nuclear medium 
  decrease when the in-medium modifications of the nucleon weak
  and electromagnetic form factors are taken into account.
  This reduction results in the enhancement of the neutrino mean free path, 
  in particular at the baryon density of around a few times of the normal nuclear matter density.}

\kword{neutrino interactions, electroweak interactions, nucleon form factors,
  differential cross section, mean free path }

\begin{document}

\maketitle

%----------------------
\section{Introduction}
\label{intro}

The majority of neutrinos in the universe are produced in the core 
collapse supernova explosion. 
The final stage of the explosion creates a hot dense proto-neutron star, which 
emits bursts of neutrinos~\cite{BL86}. 
Then the produced neutrinos propagate through the neutron star and affect 
the evolution of neutron stars.
Inside the neutron star, neutrinos scatter with the constituents of matter, mostly 
neutrons and protons, and this process determines the propagation of neutrinos, namely,
the neutrino mean free path (NMFP).
The NMFP is an important input in simulations of neutron star evolution 
as well as those of compact stars.

The electromagnetic form factors of nucleons reflect their internal structure.
Recent experimental observations in electron-nucleus scatterings suggest the in-medium 
modifications of the nucleon electromagnetic (EM) 
form factors.  
There are several issues related with the interpretation of the experimental observations in connection 
with the in-medium effects, nucleon correlations, and so on.
More detailed discussions can be found, for example, in Refs.~\cite{JLab-A-00,E03-104-11}.
In the present article, following the point of view that the properties of the quark and gluon 
substructure of nucleons change in nuclear medium and
can be estimated by effective theories of quantum chromodynamic (QCD).
We reported the effects of in-medium modified weak and EM form factors 
of the nucleon on the NMFP in dense nuclear medium.
This report is based on the recent article~\cite{HOT18}.

%--------------------------------------------------------------
\section{Neutrino scatterings with constituents of matter}
\label{scattering}

Before discussing the in-medium modifications of the nucleon weak and EM form 
factors, we briefly discuss the free space neutrino scatterings with constituents of the matter.
With the effective Lagrangian of Ref.~\cite{HOT18}, the differential cross section per volume 
of the neutrino scattering with a target particle can be calculated as
\begin{eqnarray}
\label{eq:model15}
\left( \frac{1}{V} \frac{d^3 \sigma}{d^2 \Omega^{'} dE_\nu^{'}} \right) 
&=& - \frac{1}{16 \pi^2} \frac{E_\nu^{'}}{E_\nu} 	\left[ \left( 
  \frac{G_F}{\sqrt{2}}\right)^2 \left( L_{\nu}^{\alpha\beta} \Pi^{\rm Im}_{\alpha\beta} \right)^{(\rm W)}  
  %\nonumber \\ && \mbox{}
  + \left( \frac{4\pi \alpha_{\rm em}}{q^2} \right)^2 \left( L_{\nu}^{\alpha\beta} 
  \Pi_{\alpha\beta}^{\rm Im} \right)^{(\rm EM)} 
  %\nonumber \\ && \mbox{}
  + \frac{8 \pi \,G_F \alpha_{\rm em}}{q^2 \sqrt{2}} \left( L_{\nu}^{\alpha\beta} 
  \Pi_{\alpha\beta}^{\rm Im} \right)^{(\rm INT)} 	\right], 
\nonumber \\
\end{eqnarray}
where $E'_\nu$ ($E_{\nu}$) is the final (initial) energy of the neutrino. 
For the details on the analytic formulas of the  polarization tensors for the weak and 
EM interactions and all the corresponding quantities in Eq.~(\ref{eq:model15}),  
we refer to Refs.~\cite{SWHM06,SHM05b}.
The inverse mean free path of the neutrino is straightforwardly obtained 
by integrating the differential cross section of Eq.~(\ref{eq:model15}) over 
the energy transfer $q_0^{}$ and the three-momentum transfer $\mid\vec{q}\mid$. 
The final expression for the NMFP as a function of the initial energy at a fixed baryon 
density can be obtained as~\cite{HOT18,RPL97}
\begin{align}
\label{eq:model23}
\frac{1}{\lambda (E_\nu)} &= \int_{q_0^{}}^{2E_\nu - q_0^{}} d |\vec{q}|
\int_0^{2E_\nu} d q_0^{} \frac{|\vec{q}|}{E'_\nu E_\nu} \frac{2\pi}{V} 
\frac{d^3 \sigma}{d^2 \Omega' dE'_\nu},
\end{align}
where the final and initial neutrino energies are related as $E'_\nu = E_\nu + q_0^{}$.
More detailed explanations for the determination of the lower and upper limits of the integral 
can be found in Ref.~\cite{RPL97}.

%-----------------------------------------------------------------
\section{\boldmath Models for matter} \label{mattermodels}

Following Ref.~\cite{FST96a}, we write the interactions of the nucleon in matter
as described by an effective chiral Lagrangian $\mathcal{L} = \mathcal{L}_{N} + \mathcal{L}_{M}$.
Here the Lagrangian $\mathcal{L}_{N}$ of the nucleon part is given by
\begin{eqnarray}
  \label{eq:rmf2}
  \mathcal{L}_N &=& \bar{\psi} \Bigl[ i \gamma^\mu \left( \partial_\mu^{} 
    + i v_\mu^{} + i g_\rho^{} \rho_\mu^{} + i g_\omega^{} \omega_\mu^{} \right) 
    %\nonumber \\ && \mbox{} \quad
    + g_A^{} \gamma^\mu \gamma_5 a_\mu^{} - M_N^{} + g_\sigma^{} \sigma \Bigr] \psi ,
\end{eqnarray}
where $\psi$ is the nucleon isodoublet field defined as $ \psi = \begin{pmatrix} p \\ n \end{pmatrix}$
with $M_N^{}$ being the nucleon mass. The Lagrangian $\mathcal{L}_{M}$ of the mesonic part reads~\cite{FST96a}
\begin{eqnarray}
  \label{eq:rmf4}
  \mathcal{L}_{M} &=& \frac{f_\pi^2}{4} \,\mbox{Tr} \left( \partial_\mu U \partial^\mu U^\dagger \right) 
  + \frac{f_\pi^2 m_\pi^2}{4} \, \mbox{Tr} \left( U + U^\dagger -2 \right) 
  + \frac{1}{2} \partial_\mu^{} \sigma \partial^\mu \sigma 
  - \frac{1}{2} \,\mbox{Tr} \left(\rho_{\mu \nu}^{} \rho^{\mu \nu} \right) 
  - \frac{1}{4} \omega_{\mu \nu} \omega^{\mu \nu} 
  \nonumber \\ %&& \mbox{}
  &+& \frac{1}{2} m_\omega^2 \, \omega_\mu \omega^\mu 
  + m_\rho^2 \, \mbox{Tr} \left( \rho_{\mu}^{} \rho^{\mu} \right) 
  - \frac{b}{3} M_N \left( g_\sigma \sigma \right)^3 - \frac{c}{4} \left( g_\sigma \sigma \right)^4,
\end{eqnarray}
where $\omega_{\mu\nu}^{} = \partial_\mu^{} \omega_\nu^{} - \partial_\nu^{} \omega_\mu^{}$, and 
$\rho_{\mu\nu}^{} = \partial_\mu^{} \rho_\nu^{} - \partial_\nu^{} \rho_\mu^{}$.
In the Hartree mean field approximation, the $\pi$ meson makes no contribution because of 
its negative intrinsic parity. 
Throughout the present calculation, we use $M_N^{} = 939$~MeV, $m_\rho^{} = 770$~MeV, 
$m_\omega^{} = 783$~MeV,
and $m_\sigma^{} = 520$~MeV.
We use the coupling constants determined in Ref.~\cite{GM91}, i.e.,
$\left( g_\sigma^{} / m_\sigma^{} \right)^2 = 9.148~\mbox{fm}^2$,
$\left( g_\omega^{} / m_\omega^{} \right)^2 = 4.820~\mbox{fm}^2$,
$\left( g_\rho^{} / m_\rho^{} \right)^2 = 4.791~\mbox{fm}^2$, $b = 3.478 \times 10^{-3}$, 
and $c = 1.328 \times 10^{-2}$.
(We note that similar approaches were used successfully in Ref.~\cite{SHM05b}.)

%---------------------------------------------------------------------------------
\section{Medium modifications of the nucleon form factors 
\label{medium}}

\begin{figure}[h]
\centering\includegraphics[width=0.65\columnwidth]{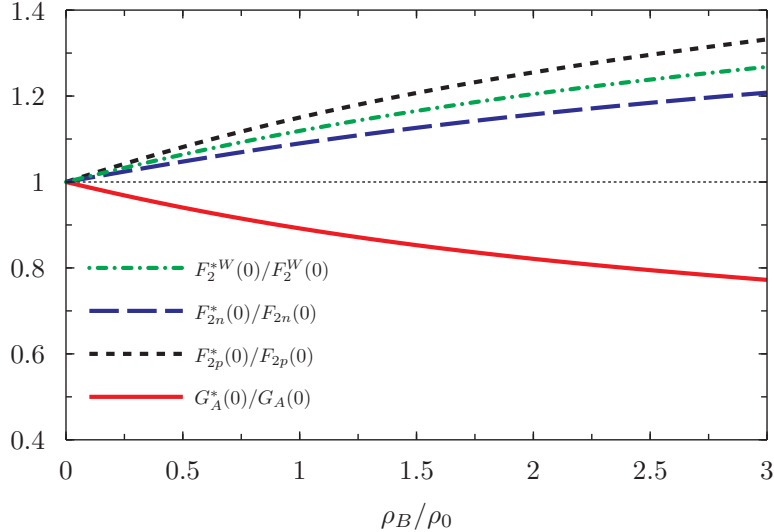}
\caption{\label{fig1} 
Ratios of the in-medium to free-space nucleon weak and EM form 
factors at the four-momentum transfer squared $q^2 = 0$
versus $\rho_B^{}/ \rho_0^{}$ with $\rho_0 = 0.15~\mbox{fm}^{-3}$.}
\end{figure}

\begin{figure}[t]
\centering\includegraphics[width=0.95\textwidth]{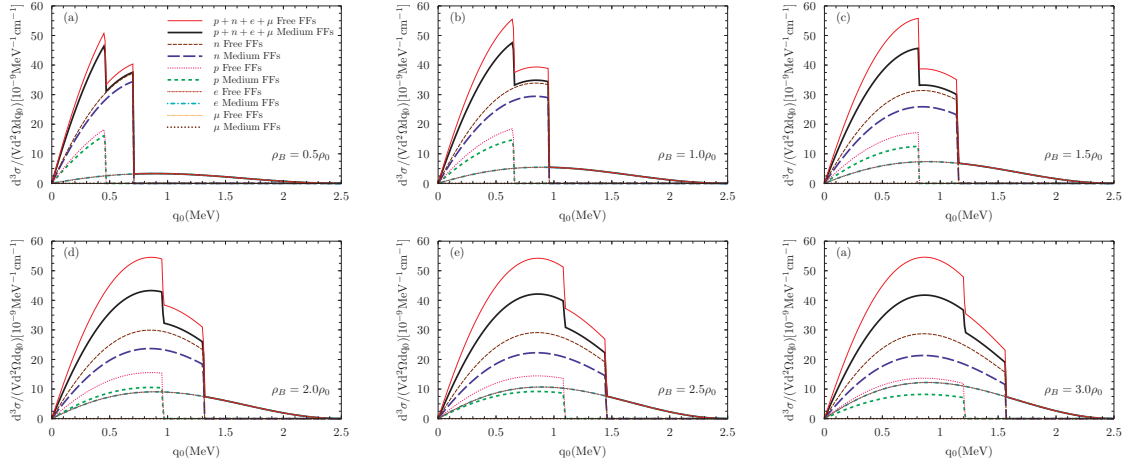}
\caption{ \label{fig2}
Differential cross sections of neutrino scatterings with the constituents of matter
as functions of the energy transfer $q_0^{}$ at the three-momentum transfer $\mid\vec{q}\mid= 2.5~\mbox{MeV}$,  
and $E_\nu = 5~\mbox{MeV}$ for (a) $\rho_B^{} = 0.5\, \rho_0^{}$, (b) $\rho_B^{} = 1.0\, \rho_0^{}$,
(c) $\rho_B^{} = 1.5\, \rho_0^{}$, (d) $\rho_B^{} = 2.0\, \rho_0^{}$,
(e) $\rho_B^{} = 2.5\, \rho_0^{}$, and (f) $\rho_B^{} = 3.0\, \rho_0^{}$.
The thick-solid, thick-long-dashed, thick-short-dashed, thick-dashed-dotted and thick-dotted 
lines are respectively the differential cross sections for total $p+n+e+\mu$, 
neutron $n$, proton $p$, electron $e$, and muon $\mu$ obtained 
with the in-medium nucleon form factors, 
while the corresponding thinner-lines are those obtained with the free-space  
nucleon form factors. Note that, for the electron $e$ and muon $\mu$ cases, the lines 
are nearly degenerate for the results obtained with the in-medium nucleon form factors 
and those obtained with the free-space ones, and difficult to distinguish.}
\end{figure}

For the estimates of the in-medium nucleon form factors,
we make use of the quark-meson coupling (QMC) model~\cite{LTTWS98b}.
Referring the further details of the QMC model to Refs.~\cite{Guichon87,STT07},
the ratios of the in-medium to free-space nucleon form factors 
$G_{E,M,A}^{\rm *\, QMC} / (G_{E,M,A}^{\rm ICBM})_{\rm free}^{}$ 
are then calculated so that the in-medium nucleon form factors can be estimated, where the superscript ICBM stands for the improved cloudy bag model.
The ratios of the in-medium to free-space nucleon form factors are then obtained
as shown in Fig.~\ref{fig2} as functions of $\rho_B^{}/ \rho_0^{}$.
It would be worthwhile to emphasize again that the form factor ratios 
presented in Fig.~\ref{fig1} are calculated based on the quark substructure of nucleons.

%%%  FIG 2
%%%%%%%%%%%%%%%%%%

%%%  FIG 3
%%%%%

%----------------------------------------------------------------
\section{Numerical results \label{numerical}}

We calculate the differential cross sections of neutrino scatterings 
with the constituents of matter at zero temperature 
as functions of the energy transfer $q_0^{}$ 
at $\mid\vec{q}\mid = 2.5~\mbox{MeV}$ with the initial neutrino energy $E_\nu = 5~\mbox{MeV}$, 
which is the typical kinematics for the cooling phase of a neutron star.
Our numerical results are shown in Fig.~\ref{fig2}, which
shows the total sum of the differential cross sections 
in vacuum (thin solid lines) and those in nuclear medium (thick solid lines)  
as well as the contributions from each target to the total sum of the differential 
cross sections (see the caption of Fig.~\ref{fig2} for details). 
In the present calculation, we set the charge radius of the neutrino $R_{V,A}=~0$ 
and neutrino magnetic moment 
$\mu_\nu = 0$ in order to focus on the different role of the nucleon form factors 
in vacuum and in medium. As a result, the shape and magnitude of
the differential cross section depend on the modifications of the form factors and effective nucleon mass.
One can verify that the impact of the in-medium nucleon weak and EM form factors 
is pronounced at higher densities.
Although the nucleon weak and EM form factors at the four-momentum transfer squared $q^2 = 0$, i.e., $F_2^{ W}(0)$ and $F_2^{\rm EM}(0)$, 
respectively, are enhanced in nuclear medium, the quenched axial-vector coupling constant $G_A^*(0)$
gives a dominant contribution to reduce the cross section, which results in the enhancement of NMFP
(See Fig.~4 of Ref.~\cite{HOT18}).

%-----------------------------------------------
\section{Summary \label{summary}}

To summarize, we have revisited our previous work on the impact of
the in-medium modifications of the nucleon weak and 
electromagnetic form factors on the neutrino scattering of Ref.~\cite{HOT18}
in the calculation of differential cross sections and 
the neutrino mean free path in dense matter using the results from a relativistic mean field model. 
The in-medium nucleon form factors are estimated by the quark-meson coupling 
model that is based on the quark degrees of freedom of the nucleon and nuclear matter
enjoying successful applications to describing the hadron and nuclear properties 
in nuclear medium.

The differential cross sections of the neutrino scatterings with the constituents of cold
matter were found to slowly decrease with increasing baryon density,
which results in the increase of the neutrino mean free path. 
This feature is sensitive to the in-medium modifications of 
the nucleon weak and electromagnetic 
form factors (in particular, that of the axial-vector form factor) as well as effective nucleon mass,  
and that the effect is pronounced for higher baryon densities.

\end{document}